\documentclass[]{interact}

\usepackage{epstopdf}% To incorporate .eps illustrations using PDFLaTeX, etc.
\usepackage{subfigure}% Support for small, `sub' figures and tables
\usepackage{caption}
\usepackage{natbib}% Citation support using natbib.sty
\bibpunct[, ]{(}{)}{;}{a}{}{,}% Citation support using natbib.sty
\theoremstyle{plain}% Theorem-like structures provided by amsthm.sty
\newtheorem{theorem}{Theorem}[section]

\newtheorem{proposition}[theorem]{Proposition}

\theoremstyle{definition}

\theoremstyle{remark}
\newtheorem{remark}{Remark}

\begin{document}

%\articletype{ARTICLE TEMPLATE}

\title{
A Numerical Approach to  Sequential Multi-Hypothesis  Testing for Bernoulli Model}

\author{
\name{Andrey Novikov\thanks{CONTACT Andrey Novikov, Universidad Aut\'onoma  Metropolitana - Unidad Iztapalapa,
    Avenida Ferrocarril San Rafael Atlixco 186, col. Leyes de Reforma 1A Secci\'on, C.P. 09310, Cd. de  M\'exico, M\'exico.
. Email: an@xanum.uam.mx} }
\affil{Metropolitan Autonomous University, Mexico City, Mexico }
}

\maketitle

\begin{abstract}
In this paper we deal with the problem of sequential testing of multiple hypotheses. The main goal  is  minimizing the  expected sample size (ESS) under restrictions on the error probabilities.  

We take, as a criterion of minimization, a weighted sum of the ESS's evaluated at some points of interest in the parameter space aiming at its minimization under restrictions on the error probabilities.

We use a variant of the method of Lagrange multipliers which is based on the minimization of an auxiliary objective  function (called Lagrangian) combining the objective function with the restrictions, taken with some constants called multipliers. Subsequently, the multipliers are used to make the solution comply with the restrictions.
%This function is defined as a  weighted sum of all the test characteristics we are interested in: the error probabilities and  the ESSs evaluated at some points of interest.
%In this paper, we use a definition of the Lagrangian function involving the ESS evaluated at any finite number of  fixed parameter points (not necessarily those representing the hypotheses).
%Particular cases of the Lagrangian function are: the Bayesian risk  in the classical Bayesian setting \citep[see, for example,][] {Baum} and the Lagrangian function used by \cite{Lorden} in the Kiefer-Weiss problem.
%Depending on the points at which the ESS is evaluated this Lagrangian function may be considered

 We develop a computer-oriented method of  minimization of the Lagrangian function, that provides, depending on the specific choice of the parameter points, optimal tests in different concrete settings, like in  Bayesian, Kiefer-Weiss and other settings. 
 
To exemplify the proposed methods  for  the particular  case of sampling from a Bernoulli population we develop  a  set of computer algorithms for  designing  sequential tests that minimize the Lagrangian function and   for the numerical evaluation of test characteristics like the error probabilities and the ESS, and other related. We implement the algorithms in the R programming language. The program code is available in a public GitHub repository. 

For the  Bernoulli model, we made a series of computer evaluations related to the  optimality of sequential multi-hypothesis tests, in a particular case of three hypotheses.
A numerical comparison with the  matrix sequential probability ratio test  is carried out. 
 
A method of solution of the multi-hypothesis Kiefer-Weiss is proposed, and is applied for a particular case of three hypotheses in the Bernoulli model.
\end{abstract}

\begin{amscode}
 62L10, 62L15, 62F03, 60G40, 62M02
 \end{amscode}

\section{Introduction}

The problem of testing multiple hypotheses is one of the oldest problems in the sequential analysis.

A traditional approach to this problem is Bayesian. It  is based  on the assumption that the hypotheses come up with some probabilities
called {\em a priori}
\citep[see][among many others]{blackwell,Baum,
Tartakovsky2014}.  

Despite that the optimal Bayesian solution can be characterized on the basis of general principles like dynamic programming or the theory of optimal stopping \citep{Shiryaev, Chow}, at least theoretically, there seems to exist a strong belief that the theoretical solution is too complex to be useful for practical purposes \citep[see, for example][]{Baum,Tartakovsky2014}. An exception is the case of two simple hypotheses where the solution is given by the classical sequential probability ratio test \cite[Wald's SPRT, see ][] {waldwolfowitz}.

For these reasons, approximate solutions of the problem have been proposed. 
One of the widely used tests, due to its simplicity, is the matrix sequential probability ratio test (MSPRT) by \cite{Armitage}. \cite{Tartakovsky2014} showed that the MSPRT  is asymptotically optimal, as error probabilities go to 0. 

%\cite{Liu} studied methods of computation of performance characteristics of the MSPRT for independent but not necessarily identically distributed observations.

Another approach that has received considerable attention through the decades is the so-called Kiefer-Weiss problem, consisting in the  minimization of the maximum value of the expected sample number (ESS), over all possible parameter points \citep{Kiefer}. 
\cite{Lorden} showed that the Kiefer-Weiss problem can be reduced to the minimization of the ESS evaluated at a specific parameter point, different from the hypothesized parameter values (so-called modified Kiefer-Weiss problem), and (in essence) used the method of Lagrange multipliers to characterize the solutions to the modified Kiefer-Weiss problem.

A generalization of the Kiefer-Weiss problem to the case of multiple hypotheses has been formulated in \cite{Tartakovsky2014} (Section 5.3) and received an asymptotic treatment in Section 5.3.1, ibid.

%To support this general setting, we want to cite a very practical context \cite{eales} 

In this paper, we propose an approach to the optimal multi-hypothesis testing based on minimization  of the  weighted ESS evaluated at parameter points not necessarily coinciding with the hypothesized values, and then use the method of the Lagrange multipliers to reduce to the minimization of the Lagrangian function.  Depending on the choice of the points for evaluating the ESS in the Lagrangian function, we obtain, in particular, the Bayesian and the  Kiefer-Weiss settings,   and more.

We apply the method of \cite{novikovMultiple} and characterize the sequential tests minimizing the Lagrangian function, for any  choice of multipliers. 
For practical applications, we propose the use of numerical methods for the Lagrange minimization, the evaluation of the characteristics  (the error probabilities, the ESS, etc.), and for finding the multiplier values to comply with the restrictions on the error probabilities.

We illustrate the proposed methods in the particular case of sampling from a Bernoulli population, where we develop a complete set of computer algorithms for all the numerical tasks  described above and implement them in the R programming language. The program code is available in a public GitHub repository in \cite{multihypothesisGit}.

Using the developed software, we run a series of numerical comparisons related to optimal properties of sequential multi-hypothesis tests in the Bernoulli model.

 First, we evaluate the performance characteristics error   of the MSPRT for a particular case of three hypotheses. The MSPRT is known to be asymptotically optimal, as the error probabilities go to 0, so the evaluations we carry out give an idea of how small  the error probabilities should be in order that the asymptotic formulas for the ESS give a  reasonably good approximation to the calculated values. We use $N=4000$ which, apparently, is sufficient for good approximations of the characteristics of non-truncated MSPRTs. 

Other comparison we carry  out is also related with the MSPRT. For a number of error probability levels, we numerically find both MSPRT and the optimal Bayes test (for uniform a priori distribution) matching the given   error probabilities (up to some precision). The results show a very high efficiency of the MSPRT.

Also we propose a method for solving a multi-hypothesis version of the Kiefer-Weiss problem, and give a numerical example.

In Section 2, we adapt the results  of \cite{novikovMultiple} to the problem of minimization of weighted ESS calculated at arbitrary parameter points.
In Section 3, we derive computational formulas for the Bernoulli model.
Numerical results are presented in Section 4.
Section 5 is a brief list of the results and suggestions for further work.
\section{Optimal sequential  multi-hypothesis tests}\label{s2}
In this section, we  formulate  some settings for the  problem of optimal multi-hypothesis testing and use the general results of \cite{novikovMultiple} for characterisation of the respective optimal solutions.

We assume that independent and identically distributed (i.i.d.) observations $X_1, X_2, \dots, X_n,\ \dots$ are potentially available to the statistician on the one-by-one basis, providing us with information about the unknown distribution of the data. Let us denote it $P_\theta$, where $\theta $ is some parameter identifying the distribution in a unique manner. We are concerned with the problem of distinguishing between a finite number of simple hypotheses 
$H_1:\theta=\theta_1$, $H_2:\theta=\theta_2$,  $\dots$, $H_k:\theta=\theta_k$, $k\geq 2$.

We follow \cite{novikovMultiple} in the notation and general assumptions.

 In particular, we consider  sequential multi-hypothesis test as a pair $\langle\psi,\phi\rangle$ of a stopping rule $\psi=(\psi_1,\psi_2,\dots, \psi_n, \dots)$, and a (terminal) decision rule $\phi=(\phi_1,\phi_2,\dots, \phi_n, \dots)$.

The elements of the stopping rule $\psi_n=\psi_n(x_1,\dots,x_n)$ are  measurable functions taking values in $[0,1]$, where the value at $(x_1,\dots,x_n)$ is interpreted as 
the conditional probability, given the observations, to stop (randomization at the stopping time).

The elements of the decision rule $\phi_n=\phi_n(x_1,\dots,x_n)$ are measurable functions of observations such that $\phi_n=(\phi_n^1, \dots, \phi_n^k)$, and $\phi_n^j\geq 0$ and $\sum_{j=1}^k\phi_n^j(x_1,\dots,x_n)\equiv 1$. Given the data $(x_1,\dots,x_n)$ observed, $\phi_n^j(x_1,\dots,x_n)$ is  interpreted as a conditional probability to accept hypothesis $H_j$, $j=1,\dots, k$ (randomization at the decision time).% with values in  governing the process of statistical decision-making. In their simplest (non-randomized) form they are functions taking values in $\{0,1\}$ and $\{1,2,\dots,k\}$, respectively,  indicating where to stop and which decision to take: the process is stopped when $\psi_n(x_1,\dots,x_n)=1$ and hypothesis $H_i$ is accepted whenever  $\phi_n(x_1,\dots,x_n)=i$, $i=1,\dots, k$. If $\psi_n(x_1,\dots,x_n)=0$ the process continues to the next step by taking an additional observation.

  The sequential test starts with observing $X_1=x_1$ (stage $n=1$). At each stage $n=1,2, \dots$ the test procedure stops with probability $\psi_n(x_1,\dots, x_n)$, given that $X_1=x_1, \dots, X_n=x_n$ are observed,  and proceeds to  taking a terminal decision. If it does not stop, the test proceeds to  taking one additional observation $X_{n+1}=x_{n+1}$ and going to stage $n+1$, etc., until the process eventually stops. When the test stops at any stage $n$ (this $n$ is called stopping time), a terminal decision is taken  accepting hypothesis $H_j$ with probability $\phi_n^j(x_1, \dots, x_n)$, conditionally on $(x_1, \dots, x_n)$. Let us denote  $\tau_\psi$ the stopping time (as a random variable) generated by the described process.

  Let $$s_n^\psi=s_n^\psi(x_1,\dots,x_n)=(1-\psi_1(x_1))\dots (1-\psi_{n-1}(x_1,\dots,x_{n-1}))\psi_n(x_1,\dots,x_n)$$
  ($s_1^\psi(x_1)=\psi_1(x_1)$ by definition). 
 
 Then the expected sample size (ESS) of the test procedure is defined as
 $$
 E_\theta \tau_\psi=\sum_{n=1}^\infty nE_\theta s_n^\psi=\sum_{n=1}^\infty nE_\theta s_n^\psi(X_1,\dots, X_n),
 $$
 provided that $\sum_{n=1}^\infty E_\theta s_n^\psi=1$, - otherwize it is infinite by definition. Here and throughout the paper, $E_\theta$ is the symbol of mathematical expectation with respect to $P_\theta$. Also we use $s_n^\psi$ (without arguments) both for $s_n^\psi(x_1,x_2,\dots,x_n)$ and for $s_n^\psi(X_1,X_2,\dots,X_n)$, depending on the context.
So do we when dealing with other functions like $\psi_n$, $\phi_n$, etc.

  Other characteristics of a sequential test $\langle\psi,\phi\rangle$ are the error probabilities defined as
  \begin{equation*}
   \alpha_{ij}(\psi,\phi)=\sum_{n=1}^\infty E_{\theta_i}s_n^\psi \phi_n^j,\;1\leq i\not=j\leq k.
  \end{equation*}
  Another natural way to define  error probabilities is  less detailed:    
\begin{equation*}
   \alpha_{i}(\psi,\phi)=\sum_{n=1}^\infty E_{\theta_i}s_n^\psi(1-\phi_n^i)
=  \sum_{j:j\not =i}\alpha_{ij}(\psi,\phi),\;1\leq i\leq k.
  \end{equation*} In the case of two hypotheses the definitions are equivalent. 

For $k=2$, the classical result of \cite{waldwolfowitz} states that the sequential probability ratio test (SPRT) minimizes both $E_{\theta_1 }\tau_\psi$ and $E_{\theta_2} \tau_\psi$ in the class of sequential tests $\langle\psi,\phi\rangle$ such that
\begin{equation*}
 \alpha_1(\psi,\phi)\leq \alpha_1, \quad \alpha_2(\psi,\phi)\leq \alpha_2,
\end{equation*}
where $\alpha_1$ and $\alpha_2$ are the error probabilities of the SPRT. 

To the best of our knowledge, no direct generalizations of this result exist for $k>2$. 
For this reason, we propose  weaker settings.

Let us choose some parameter points  $\vartheta_i$, $i=1,\dots,K$  and the weights $\gamma_i$, $i=1,\dots,K$ being these non-negative numbers such that $\sum_{i=1}^K\gamma_i=1$, $K\geq 1$.
Formally, we propose to minimize the weighted ESS

\begin{equation}\label{8-1}\
 C_{\gamma,\vartheta}(\psi)=\sum_{i=1}^K\gamma_i E_{\vartheta_i}\tau_\psi                      
\end{equation}
over all sequential multi-hypothesis tests subject to
\begin{equation}\label{1}
 \alpha_{ij}(\psi,\phi)\leq \alpha_{ij},\quad 1\leq i,j\leq k, \quad i\not = j
\end{equation}
or to
\begin{equation}\label{2}
 \alpha_{i}(\psi,\phi)\leq \alpha_{i}, \quad 1\leq i\leq k
\end{equation}
where $\alpha_{ij}$ and $\alpha_i$ are some positive numbers.

To support this formulation, let us refer to a very practical context of optimal group-sequential testing in the case of two hypotheses. For testing the mean of a normal distribution with known variance, \cite{eales}  considered five settings for the ESS minimization under restrictions on the error probabilities. Four of them, namely, $F_1$ to $F_4$ \citep[see][]{eales} are of type \eqref{8-1}, with different choices of $K$, $\vartheta_i$ and $\gamma_i$. $F_5$ is also a kind of weighted ESS but of continuous type, which is quite possible to be treated by our method, but for the time being stays beyond the scope.
Generalizations of these settings to the case of more than two hypotheses and infinite horizons are straightforward.

Given that the formulated problem is a problem of a minimization under restrictions,
we want to use  the Lagrange multipliers method.
By the principle of the Lagrange method, to minimize $C_{\gamma,\vartheta} $ under restrictions \eqref{1} one should be able to minimize  the Lagrangian function 
\begin{equation}\label{3}
L(\psi,\phi)=C_{\gamma,\vartheta}(\psi)+\sum_{1\leq i\not= j\leq k}\lambda_{ij}\alpha_{ij}(\psi,\phi),
\end{equation}  
with any constant multipliers $\lambda_{ij}\geq 0$, and to find the values of the multipliers  for which equalities in \eqref{1} hold.
Respectively, the problem of minimization under conditions \eqref{2} reduces to minimization of 
\begin{equation}\label{4}
L(\psi,\phi)=C_{\gamma,\vartheta}(\psi)+\sum_{1\leq i\leq k}\lambda_{i}\alpha_{i}(\psi,\phi),
\end{equation} with multipliers $\lambda_i$, $i=1,\dots,k$, and finding the values of $\lambda_i$ for which equalities in \eqref{2} hold.
It is easy to see that \eqref{4} is a particular case of \eqref{3} with $\lambda_{ij}=\lambda_i$ for all $j=1,2,\dots,k$, $j\not=i$,
so in what follows we  focus on the  minimization of \eqref{3}.

It is not difficult to see that in the particular case when $\theta_i=\vartheta_i$, $i=1,2\dots,k=K$ the Lagrangian function \eqref{3} can be considered Bayesian risk \citep[see, for example,][among many others]{Baum} corresponding to the a priori distribution $(\gamma_1, \dots,\gamma_k)$ on the set of parameter points $\{\theta_1,\dots, \theta_k\}$, where $\lambda_{ij}/\gamma_i$ can be interpreted as conditional loss from accepting $H_j$ when $H_i$ is true. Thus, the minimization of \eqref{3} readily  solves the problem of optimal Bayesian tests for $k$ hypotheses.

The well-known modified Kiefer-Weiss  problem \citep[see, for example,][]{Lorden} also easily embeds into this scheme by taking $\gamma_1=1$, $K=1$, and $\vartheta_1$ between the hypothesized values $\theta_1$ and $\theta_2$, being $k=2$. And this gives rise to a multi-hypothesis version of the Kiefer-Weiss problem, starting from a modified version of it, with $\vartheta_1,\dots,\vartheta_{k-1}$  such that $\theta_1<\vartheta_1<\theta_2<\vartheta_2<\dots<\vartheta_{k-1}<\theta_k$ and with  some weights  $\gamma_1,\gamma_2,\dots,\gamma_{k-1}$, adding up to 1,  as additional parameters. To our knowledge, there are no known non-asymptotic solutions of the multi-hypothesis Kiefer-Weiss problem, and this could be a basis for one.
 
Now, let us  characterize the tests which minimize the Lagrangian function \eqref{3}, for a given  set of multipliers. 
 It is worth noting that $L(\psi,\phi)$ implicitly depends on the Lagrange multipliers, therefore all the constructions below will also (implicitly) depend on $\lambda_{ij}$, as well as on other elements of problem setting, like $\theta_i$ and $\vartheta_i$, etc.

 First of all, in a very standard way it can be shown that there is a universal decision rule $\phi$ that minimizes $L(\psi,\phi)$ whatever fixed $\psi$ \citep[see][]{novikovMultiple}.

Let us assume that $P_\theta$ is absolutely continuous with respect to a $\sigma$-finite measure $\mu$ and denote $f_\theta$ its Radon-Nikodym derivative. Also denote
$f_\theta^n=f_\theta^n(x_1,\dots, x_n)=\prod_{i=1}^nf_\theta(x_i)$, and let $f^n_{{\gamma\vartheta}}=\sum_{i=1}^K \gamma_if_{\vartheta_i}^n$.
Define \begin{equation}
  v_n=\min_{1\leq j\leq k}\sum_{i:i\not =j}\lambda_{ij}f_{ \theta_i}^n.
 \end{equation}
Let a decision rule $\phi$ be such that 
 \begin{equation}\label{2-9}
\phi_{n}^j=0 \quad\mbox{whenever}\quad \sum_{i:i\not
=j}\lambda_{ij}f_{\theta_i}^n>v_n
\end{equation}
(in the case of equality in \eqref{2-9} $\phi_{n}^j$ can be arbitrarily randomized between those $j$  sharing this equality, with the only requirement that $\sum_{j=1}^k \phi_{n}^j\equiv 1$).
It follows from Theorem 3 in \cite{novikovMultiple} that 
\begin{equation}\label{2-1}
L(\psi)=\inf_{\phi}L(\psi,\phi)=\sum_{n=1}^\infty \int s_n^\psi\left(nf_{{{\gamma\vartheta}}}^{n}+v_n
\right) d\mu^n,
\end{equation}
and we have an optimal stopping problem of minimizing \eqref{2-1} over stopping rules $\psi$.

The problem is first solved in the class of truncated tests, i.e. those not taking more than a finite number $N$ of observations. Let $\mathcal S^N$ be the set of all such stopping rules that $(1-\psi_1)\dots(1-\psi_N)\equiv 0$.

Let us define operator $\mathcal I_n$ in the following way.
For any measurable non-negative $v=v(x_1, \dots, x_n)$ let $$\mathcal I_nv=(\mathcal I_nv)(x_1,\dots,x_{n-1})=\int v(x_1,\dots,x_{n}) d\mu(x_n).$$
Now, starting from 
\begin{equation*}\label{2-3}
V_N^N\equiv v_N,
\end{equation*}
define recursively over $n=N,N-1,\dots, 2$
\begin{equation*}\label{2-4}
V_{n-1}^N=\min\{v_{n-1},f_{{\gamma\vartheta}}^{n-1}+\mathcal I_nV_n^N\}.
\end{equation*}
Then for any $\psi\in\mathcal S^N$
\begin{equation}\label{2-5}
L(\psi)\geq 1+\mathcal I_1 V_1^N,
\end{equation}
and there is an equality in \eqref{2-5} if for all $n=1,2,\dots, N-1$
\begin{equation}\label{2-6}
\psi_n= I_{\{v_{n}\leq f_{{\gamma\vartheta}}^{n}+\mathcal I_{n+1}V_{n+1}^N\}},
\end{equation}
where $I_A$ denotes the indicator function of the event $A$.
In this way, stopping rule $\psi $ in \eqref{2-6} minimizes $L(\psi)$ in the class of truncated stopping rules $\mathcal S^N$. 
Any $\psi_n$ may be arbitrarily randomized between samples $(x_1, \dots, x_n)$ for which
there is an equality in the inequality under the indicator function in \eqref{2-6}. This gives the same value of $L(\psi)$. The details can be found in \cite{novikovMultiple}.

The optimal non-truncated tests can be found passing to the limit, as $N\to \infty$, provided that \begin{equation}\label{9-1}
 \int v_n d\mu^n\to 0, \quad\mbox{as}\quad n\to\infty,
\end{equation}
\citep[see Remark 7 in][]{novikovMultiple}.
In the case of i.i.d. observations we are considering in this paper, \eqref{9-1} holds without any additional conditions. The formal proof of this fact can be found in the Appendix.

The construction of the optimal non-truncated test is as follows.
First of all, it is easy to see that $V_n^{N+1}\leq V_n^N$, so there exists $V_n=\lim_{N\to\infty} V_n^N$, $n=1,2,\dots$
Then it follows from \eqref{2-5} that
\begin{equation}\label{2-7}
L(\psi)\geq 1+\mathcal I_1 V_1,
\end{equation}
and the right-hand side in \eqref{2-7} is attained if 
\begin{equation}\label{2-8}
\psi_n=I_{\{v_{n}\leq f_{{\gamma\vartheta}}^{n}+\mathcal I_{n+1}V_{n+1}\}}
\end{equation}
for all $n=1,2,\dots$
In this way, we obtain tests $\langle\psi,\phi\rangle$ 
with $\psi$ satisfying \eqref{2-8} and $\phi$ satisfying \eqref{2-9}
which minimize the Lagrangian function $L(\psi,\phi)$. 

We propose using  numerical methods for construction of the truncated  tests minimizing the Lagrangian function. For the Bernoulli model, we develop numerical algorithms for this and implement them in the form of a computer program in the R programming language. 
Having the means for minimizing the Lagrangian function, to obtain optimal sequential tests in the conditional setting (i.e. those minimizing $C_{\gamma,\vartheta}$ under conditions \eqref{1}) we  need to find Lagrangian multipliers $\lambda_{ij}$, $1\leq i\not=j\leq k$, providing a test  \eqref{2-9}-\eqref{2-6} for which  
equalities in \eqref{1} hold. Respectively, the minimization of $C_{\gamma,\vartheta}$ under conditions \eqref{2}  reduces to finding $\lambda_i$, $i=1,\dots, k$  such that for the test in \eqref{2-9}-\eqref{2-6}, with  $\lambda_{ij}=\lambda_i$  for  $1\leq j\not = i\leq k$, for which there are all  equalities in \eqref{2}.

In no way can one be sure that  such $\lambda_{ij}$ exist for every combination of $\alpha_{ij}$ (not even in  the classical case of two hypotheses). On the other hand, {\em every} combination of  $\lambda_{ij}$ employed in \eqref{2-9}-\eqref{2-6}, produces an optimal test $\langle\psi,\phi\rangle$ in the conditional setting,  if one takes its error probabilities  as $\alpha_{ij}$ in \eqref{1}
(i.e. $\alpha_{ij}=\alpha_{ij}(\psi,\phi)$)  (or, respectively, as $\alpha_i$ in \eqref{2}, that is $ \alpha_i=\alpha_{i}(\psi,\phi)$).

Having at hand a computer program for the Lagrange minimization, finding the multipliers providing a tolerable level of the error probabilities is a question of some  trial-and-error look-ups, because larger values of $\lambda_{ij}$ make $\alpha_{ij}$ smaller, {\em grosso modo}. As an alternative, general-purpose computer algorithms of numerical optimization can be used to get as close as possible to the desired values of $\alpha_{ij}$ by moving the input values of $\lambda_{ij}$, for example, the method of \cite{neldermeadarticle}.

For the non-truncated tests, we propose  using approximations by truncated tests. We illustrate all this technique  on the particular case  of Bernoulli distribution in the subsequent sections.
\section{Optimal sequential tests for sampling from a Bernoulli population}

In this section, we apply the general results of Section \ref{s2}
to the model of Bernoulli observations. In this way we obtain a complete set of computer algorithms for computing the tests that minimize the Lagrangian function, and their numerical characteristics, in the Bernoulli model. For the determination of the values of the Lagrange multipliers general-purpose computer algorithms will be used.
%According to the problem setting of Section 2, ``optimal'' will be a short name for the tests minimizing the Lagrangian function, if not  stated otherwize. 

\subsection{Construction of optimal tests}

We apply the results of Section \ref{s2} to the model of sampling from a Bernoulli population, in which  case $f_\theta(x)=\theta^x(1-\theta)^{1-x}$, $x=0,1$, and $f_\theta^n(x_1,\dots,x_n)=\theta^{s_n}(1-\theta)^{n-s_n}$ with $s_n=\sum_{i=1}^nx_i$.

Let
\begin{equation*}
 g_\theta^n(s)={n\choose s}\theta^{s}(1-\theta)^{n-s},\; 0\leq s\leq n
\end{equation*}
be the probability mass function corresponding to the sufficient statistic $S_n=\sum_{i=1}^n X_i$ (binomial distribution with parameters $n$ and $\theta$).
Define \begin{equation}\label{3-1}
  u_n=u_n(s)=\min_{1\leq j\leq k}\sum_{i:i\not =j}\lambda_{ij}g_{ \theta_i}^n(s),\; 0\leq s\leq n,
 \end{equation}
 and let
 $$
 g_{\gamma\vartheta}^n(s)=\sum_{i=1}^K \gamma_i g_{\vartheta_i}^n(s),\; 0\leq s\leq n 
 $$
 Let us define the operator $\mathcal J_n$ defined for any  function $U(s)$, $0\leq s\leq n$, as
 \begin{equation}\label{6-4}
  \mathcal J_{n} U(s)=
U(s)\frac{n-s}{n}+U(s+1)\frac{s+1}{n},\; 0\leq s\leq n-1,
\end{equation}
for $n=2,3,\dots$
 Starting from
 \begin{equation*}\label{2-10}
U_N^N(s)=u_N(s),\; 0\leq s\leq N, 
 \end{equation*}
define recursively for $n=N-1, N-2,\dots, 1$
%   \begin{equation}\label{2-11}
%   U_{n}^N(s)=
% \min\left\{u_{n}(s),g_{{\gamma\vartheta}}^{n}(s)+U_{n+1}^N(s)\frac{n+1-s}{n+1}+U_{n+1}^N(s+1)\frac{s+1}{n+1} \right\},\; 0\leq s\leq n.
% \end{equation}

  \begin{equation}\label{2-11}
  U_{n}^N(s)=
\min\left\{u_{n}(s),g_{{\gamma\vartheta}}^{n}(s)+\mathcal J_{n+1}U_{n+1}^N(s) \right\},\; 0\leq s\leq n.
\end{equation}

\begin{proposition}\label{p1}
 For $m=1,2\dots,N-1$
\begin{equation}\label{6-5}
  \mathcal J_{m+1}U_{m+1}^N(s_{m})={m\choose s_{m}}\mathcal I_{m+1}V^N_{m+1}(x_1,\dots,x_{m})
 \end{equation}
 where $s_m=\sum_{i=1}^m x_i$.
\end{proposition}
{\bf Proof.} By induction over $m=N-1,N-2,\dots,1$. 
For $m=N-1$ we have
$$  \mathcal J_{m+1}U_{m+1}^N(s_{m})= \mathcal J_{N}U_{N}^N(s_{N-1})=
\mathcal J_{N}u_{N}(s_{N-1})
$$
$$
=u_N(s_{N-1})\frac{N-s_{N-1}}{N}+u_N(s_{N-1}+1)\frac{s_{N-1}+1}{N}$$
$$=
{N\choose s_{N-1}}v_N(x_1,\dots,x_{N}=0)\frac{N-s_{N-1}}{N}+{N\choose s_{N-1}+1}v_N(x_1,\dots,x_{N}=1)\frac{s_{N-1}+1}{N}
$$
$$
=
{N-1\choose s_{N-1}}\mathcal I_{N}V^N_{N}(x_1,\dots,x_{N-1})= {m\choose s_{m}}\mathcal I_{m+1}V^N_{m+1}(x_1,\dots,x_{m})
$$
Let us suppose now that \eqref{6-5} holds for some $m\leq n\leq N-1$. Then for $m=n-1$
$$  \mathcal J_{m+1}U_{m+1}^N(s_{m})= \mathcal J_{n}U_{n}^N(s_{n-1})$$
$$=
U_n^N
(s_{n-1})\frac{n-s_{n-1}}{n}+U_n^N(s_{n-1}+1)\frac{s_{n-1}+1}{n}$$
$$=
{n\choose s_{n-1}}V_n^N(x_1,\dots,x_{n}=0)\frac{n-s_{n-1}}{n}+{n\choose s_{n-1}+1}V_n^N(x_1,\dots,x_{n}=1)\frac{s_{n-1}+1}{n}$$
$$
=
{n-1\choose s_{n-1}}\mathcal I_{n}V^N_{n}(x_1,\dots,x_{n-1})= {m\choose s_{m}}\mathcal I_{m+1}V^N_{m+1}(x_1,\dots,x_{m})
$$
$\Box$

It is easy to see  that the optimal decision rule \eqref{2-9}  can be expressed in terms of the sufficient statistic $s_n$:
  \begin{equation}\label{3-4}
\phi_{n}^j(s_n)=0 \quad \text{whenever}\quad \sum_{i:i\not
=j}\lambda_{ij}g_{\theta_i}^n(s_n)>u_n(s_n),
\end{equation}
and it follows from Proposition \ref{p1}  that the optimal truncated stopping rule \eqref{2-6} as well:
\begin{equation}\label{3-5}
\psi_n(s_n)=I_{\{u_{n}\leq g_{{\gamma\vartheta}}^{n}+\mathcal J_{n+1}U_{n+1}^N\}}(s_n),
\end{equation}
for $n=1,2\dots,N-1$, and the optimal non-truncated one as
\begin{equation}\label{3-6}
\psi_n(s_n)=I_{\{u_{n}\leq g_{{\gamma\vartheta}}^{n}+\mathcal J_{n+1}U_{n+1}\}}(s_n)
\end{equation}
with $U_n=\lim_{N\to\infty}U_n^N$ for all $n=1,2,\dots$

Formulas \eqref{3-4}-\eqref{3-5} provide  a truncated test which has an {\em exact} optimality property (neither asymptotic nor approximate), whatever be $k\geq 2$, $\theta_1,\dots,\theta_k,\gamma_1,\dots,\gamma_K, \vartheta_1,\dots,\vartheta_K, K\geq 1$, $N\geq 2$ and Largange multipliers $\lambda_{ij}\geq 0$, $1\leq i\not =j\leq k$. 

Furthermore, they suggest a computational algorithm for evaluating the elements of optimal sequential test: start from step $N$ calculating $\phi_N$ for all $0\leq s \leq N$ (which is based on weighted sums of binomial probabilities with parameters $N$ and $\theta_i$, $i=1,2,\dots,k$, according to \eqref{3-4}),
and recurrently use  \eqref{2-11} for steps $n=N-1,N-2,\dots,1$ to calculate $U_n^N(s)$ for all $0\leq s\leq n$, marking those $s$ for which
$$
u_n(s)>g_{{\gamma\vartheta}}^n(s)+\mathcal J_{n+1}U_{n+1}^N(s)
$$
as belonging to the continuation region (by virtue of \eqref{3-5}); for all other $s$ storing the terminal decision   based on \eqref{3-4} as that corresponding to $s$.

We implemented this algorithm in the form of a function in the R programming  language \citep{R}; the source code is  available in a public GitHub repository in \cite{multihypothesisGit}.  The documentation can be found in the repository.

Making $N$ large enough we can approximate the optimal non-truncated test corresponding to \eqref{3-6}. In particular, this can be helpful
when the optimal infinite-horizon test is in fact truncated. This happens, for example, in the case of modified Kiefer-Weiss problem, corresponding (in our notation) to the case of two hypotheses with $\theta_1<\vartheta_1<\theta_2$, $k=2$, $K=1$ \citep[see][]{Lorden}. Below in Section 4 we give another example of this possibility, in a multi-hypothesis context.

Despite that the test obtained in this subsection does not have a closed form (instead, {\em all} the values of the optimal rules \eqref{3-4} -- \eqref{3-5} are stored in the computer memory), we believe it can be quite practical for many applications which do not require more than some thousands of steps. If they do, one could try the algorithm with a maximum  number of steps their computer will withstand, to see if the performance requirements could be met with that reduced number of steps. If not, more computer power might be needed. 

\subsection{Evaluation of performance characteristics}
\label{s3.2}
We derive in this part  computational formulas for performance characteristics of sequential multi-hypothesis tests  for the Bernoulli model.

Let $\langle\psi,\phi\rangle$ be any sequential  multi-hypothesis test based on sufficient statistics: $\psi_n=\psi_n(s_n)$, $\phi_n=\phi_n(s_n)$ with $\psi\in \mathcal S^N$. The test $\langle\psi,\phi\rangle$ is arbitrary but will be held fixed throughout this subsection, so it will be suppressed in the notation.

\begin{proposition} Define \begin{equation}\label{25a_1}a_j^N(s;\theta)=g_\theta^N(s)\phi_N^j(s),\; s=0,1,\dots,N,\;  j=1,2,\dots,k,\end{equation} 
and, recursively over $n=N-1,N-2,\dots,1$, 
\begin{eqnarray*}\label{25a_2}a_j^{n}(s;\theta)&=&g_\theta^{n}(s)\psi_{n}(s)\phi_n^j(s)\nonumber\\
&&+\left(a_j^{n+1}(s;\theta)\frac{n+1-s}{n+1}+a_j^{n+1}(s+1;\theta)\frac{s+1}{n+1}\right)(1-\psi_n(s)),\end{eqnarray*}
$s=0,1,\dots,n$, $ j=1,\dots k$.

Then the probability to accept hypothesis $H_j$, given that the true parameter
is $\theta$, can be  calculated as  $a_{j}^0(\theta)=a_{j}^1(0;\theta)+a_{j}^1(1;\theta)$. In particular, $\alpha_{ij}(\psi,\phi)=a_{j}^0(\theta_i)$, $i\not=j$.
\end{proposition}
{\bf Proof.}
Let us denote $A_j^n=A_j^n(\psi,\phi)$ the event meaning that hypothesis $H_j$ is accepted at or after step $n$ (following the rules of the test $\langle\psi,\phi\rangle$), $n=1,2,\dots,N$.

Let us  first prove, by induction over $n=N,N-1,\dots,1$, that
\begin{equation}\label{5-1}
a_j^n(S_n;\theta)=P_\theta(A_j^n|X_1,\dots,X_n)g_\theta^n(S_n)
\end{equation}
For $n=N$, \eqref{5-1} follows from \eqref{25a_1} and the definition of the decision rule $\phi$.

Let us suppose now that \eqref{5-1} holds for some $n\leq N$. Then
\begin{eqnarray}\label{5-2}
&&a_j^{n-1}(S_{n-1};\theta)=g_\theta^{n-1}(S_{n-1})\psi_{n-1}(S_{n-1})\phi_{n-1}^j(S_{n-1})\nonumber\\
&& \quad+\Big[a_j^{n}(S_{n-1};\theta)\frac{n-S_{n-1}}{n}+a_j^{n}(S_{n-1}+1;\theta)\frac{S_{n-1}+1}{n}\Big](1-\psi_{n-1}(S_{n-1})).
\end{eqnarray}
But, by the supposition,
\begin{eqnarray}
 \;& &a_j^{n}(S_{n-1};\theta)\frac{n-S_{n-1}}{n}+a_j^{n}(S_{n-1}+1;\theta)\frac{S_{n-1}+1}{n}\nonumber\\
 &\;&=P_\theta(A_j^n|X_1,\dots,X_{n-1},X_n=0)g_\theta^n(S_{n-1})\frac{n-S_{n-1}}{n}\nonumber\\
 &&\quad+ P_\theta(A_j^n|X_1,\dots,X_{n-1},X_n=1)g_\theta^n(S_{n-1}+1)\frac{S_{n-1}+1}{n}\nonumber\\
 &\;&=(P_\theta(A_j^n|X_1,\dots,X_{n-1},X_n=0)(1-\theta)\nonumber\\
 &&\quad+ P_\theta(A_j^n|X_1,\dots,X_{n-1},X_n=1)\theta )g_\theta^{n-1}(S_{n-1})\nonumber\\
&\;& =
  P_\theta({A_j^{n}}|X_1,\dots,X_{n-1})g_\theta^{n-1}(S_{n-1})\nonumber
 \end{eqnarray}
Therefore, \eqref{5-2} equals
\begin{eqnarray}
 &&\left(\psi_{n-1}\phi_{n-1}^j+P_\theta({A_j^{n}}|X_1,\dots,X_{n-1})(1-\psi_{n-1})\right)g_\theta^{n-1}(S_{n-1})\nonumber\\
&\;&=P_\theta(A_{j}^{n-1}|X_1,\dots,X_{n-1})g_\theta^{n-1}(S_{n-1}).\nonumber
 \end{eqnarray}
Now that \eqref{5-1} is proved, we apply it for $n=1$
and have
\begin{equation*}\label{5-3}
a_j^1(1;\theta)=P_\theta(A_j^1|X_1=1)\theta\quad\text{and}\quad a_j^1(0;\theta)=P_\theta(A_j^1|X_1=0)(1-\theta),
\end{equation*}
thus,
\begin{equation*}
 a_j^1(0;\theta)+a_j^1(1;\theta)=P_\theta(A_j^1|X_1=1)\theta+P_\theta(A_j^1|X_1=0)(1-\theta)=P_\theta(A_j^1)=a_j^0(\theta).
\end{equation*}

$\Box$

In an analogous way, characteristics of sample number can be treated.
\begin{proposition}\label{p2} For any stopping rule $\psi$ 
define for any $m\geq 1$ 
\begin{equation}\label{4-1}
                       b_{m}^m(s;\theta)=g_\theta^m(s)(1-\psi_m(s)),\; s=0,1,\dots,m,
                      \end{equation}
 and, recursively over $n=m-1,m-2,\dots,1$, 
\begin{equation}\label{4-3}
b_n^{m}(s;\theta)=\left(b_{n+1}^{m}(s;\theta)\frac{n+1-s}{n+1}+b_{n+1}^{m}(s+1;\theta)\frac{s+1}{n+1}\right)(1-\psi_{n}(s)),
\end{equation}
 $s=0,1,\dots,n$.
Then  $P_\theta(\tau_\psi>m)=b_1^m(0;\theta)+b_1^m(1;\theta)$.
\end{proposition}
{\bf Proof.}
Let us denote $B_n^m=B_n^m(\psi)$, $n=1,2,\dots,m$, the event meaning that the test following  the stopping rule  $\psi$ does not stop at any step between $n$ and $m$, inclusively.

Let us first prove, by induction over $n=m,m-1,\dots,1$, that
\begin{equation}\label{6-1}
b_n^m(S_n;\theta)=P_\theta(B_n^m|X_1,\dots,X_n)g_\theta^n(S_n)
\end{equation}
For $n=m$, \eqref{6-1} follows from \eqref{4-1}.
Let us suppose now that \eqref{6-1} holds for some $n\leq m$. Then
\begin{eqnarray*}\label{6-2}
&&b_{n-1}^{m}(S_{n-1};\theta)=
\left(b_n^{m}(S_{n-1};\theta)\frac{n-S_{n-1}}{n}+b_n^{m}(S_{n-1}+1;\theta)\frac{S_{n-1}+1}{n}\right)(1-\psi_{n-1})\nonumber\\
& &\quad=\Big[P_\theta(B_n^m|X_1,\dots,X_{n-1},X_n=0)g_\theta^n(S_{n-1})\frac{n-S_{n-1}}{n}\nonumber\\
 &&\quad\quad+\; P_\theta(B_n^m|X_1,\dots,X_{n-1},X_n=1)g_\theta^n(S_{n-1}+1)\frac{S_{n-1}+1}{n}\Big](1-\psi_{n-1})\nonumber\\
 &&\quad=\Big[P_\theta(B_n^m|X_1,\dots,X_{n-1},X_n=0)(1-\theta)\nonumber\\
&&\quad\quad+\; P_\theta(B_n^m|X_1,\dots,X_{n-1},X_n=1)\theta\Big](1-\psi_{n-1})g_\theta^{n-1}(S_{n-1})\nonumber\\
&&\quad =
  P_\theta( B_n^{m}(1-\psi_{n-1})|X_1,\dots,X_{n-1})g_\theta^{n-1}(S_{n-1})\nonumber\\
&&\quad 
=
  P_\theta( B_{n-1}^{m}|X_1,\dots,X_{n-1})g_\theta^{n-1}(S_{n-1})\nonumber
 \end{eqnarray*}
Now that \eqref{6-1} is proved, we apply it for $n=1$
and have
\begin{equation*}\label{6-3}
b_1^m(1;\theta)=P_\theta\{B_1^m|X_1=1\}\theta\quad\text{and}\quad b_1^m(0;\theta)=P_\theta\{B_1^m|X_1=0\}(1-\theta),
\end{equation*}
thus,
\begin{equation*}
 b_1^m(0;\theta)+b_1^m(1;\theta)=P_\theta(B_1^m|X_1=1)\theta+P_\theta(B_1^m|X_1=0)(1-\theta)=P_\theta(B_1^m)=P_\theta(\tau_\psi>m).
\end{equation*}
$\Box$

It follows from Proposition \ref{p2} that if $\psi\in \mathcal S^N$, then
\begin{equation}\label{4-2}
E_\theta\tau_\psi=\sum_{m=1}^NP_\theta(\tau_\psi\geq m)=1+\sum_{m=1}^{N-1}(b_1^m(0;\theta)+b_1^m(1;\theta)).
\end{equation}

If a stopping rule $\psi$ is not truncated, we can use \eqref{4-2} to approximate
$E_\theta \tau_\psi$, noting that $E_\theta\min\{\tau_\psi,N\}\to E_\theta \tau_\psi$, as $N\to\infty$, by the theorem of monotone convergence, and $\min\{\tau_\psi,N\}$ corresponds to the truncated rule $\psi^N=(\psi_1,\dots, \psi_{N-1},1, \dots)\in \mathcal S^N$. Applying \eqref{4-2} to $\psi^N$
we see that
$E_\theta\min\{\tau_\psi,N\}=1+\sum_{m=1}^{N-1}(b_1^m(0;\theta)+b_1^m(1;\theta))$, thus 
\begin{equation*}
E_\theta \tau_\psi=1+\sum_{m=1}^{\infty}(b_1^m(0;\theta)+b_1^m(1;\theta)).
\end{equation*}

Dealing with expectations, a more direct way to evaluate \eqref{4-2} is  incorporating the summation in \eqref{4-2} into the 
inductive evaluations in \eqref{4-3}. This is done in the following
\begin{proposition} For a stopping rule $\psi$, 
define \begin{equation*}\label{4-4}
                       c_{N}^{N}(s;\theta)=g_\theta^{N}(s)(1-\psi_{N}(s)),\; s=0,1,\dots,N,
                      \end{equation*}
 and, recursively over $n=N-1,N-2,\dots,1$, 
\begin{equation*}\label{4-5}
c_n^{N}(s;\theta)=\left(g_\theta^{n}(s)+c_{n+1}^{N}(s;\theta)\frac{n+1-s}{n+1}+c_{n+1}^{N}(s+1;\theta)\frac{s+1}{n+1}\right)(1-\psi_{n}(s)),
\end{equation*}
 $s=0,1,\dots,n$.
 Then
 \begin{equation}\label{4-6}
  E_\theta  \min\{\tau_\psi, N+1\}=1+c_1^{N}(0;\theta)+c_1^{N}(1;\theta)
 \end{equation}
\end{proposition}
Again, passing to the limit in \eqref{4-6}, as $N\to\infty$, we obtain
\begin{equation*}\label{4-7}
  E_\theta \tau_\psi=1+\lim_{N\to\infty}(c_1^{N}(0;\theta)+c_1^{N}(1;\theta))
 \end{equation*}
 
 We implemented the algorithms presented in this subsection in the R programming language; the source  code is available   in \cite{multihypothesisGit}.
 
It should be noted that the algorithms for performance evaluations in this subsection are applicable to any truncated test based on sufficient statistics, and not only to the optimal test of Subsection 3.1. In particular, we included in the program implementation a function producing the structure of the (truncated version of) the  matrix sequential probability ratio test (MSPRT), enabling in this way all the performance evaluations of this subsection for the truncated MSPRT as well. Because an MSPRT
for two hypotheses is an SPRT, this also covers the performance evaluation of  truncated SPRTs. Also, an implementation of the Monte Carlo simulation for the performance evaluation is provided as a part of the program code.

%But we do not think this could be of much practical importance,
%unless , or  there is a closed-form expression of the optimal test.
%The former possibility is presentedThe latter corresponds to the classical case of Wald's SPRT for two simple hypotheses (and its extension in \cite{Burkholder}). To our knowledge, no other models are described in the literature where the optimal sequential tests have a closed form.   

%(as, for example, in the case of two hypotheses, where the optimal test is known to be the classical Wald's SPRT). 

%Because all our methods are essentially numerical, we can control the values of all numerical characteristics at any time, and say that we get close to the optimal non-truncated test when a substantial growth of $N$ leaves the calculated characteristics largely unaffected. This type of control dealing with sequential statistical procedures 8s not any uncommon. For example, \cite{Young}

 \section{Applications. Numerical results}
 
In this section we  apply the theoretical results of the preceding sections to construction and performance evaluation of sequential tests in the Bernoulli model.

\subsection{Efficiency of the MSPRT }
 In this subsection, we evaluate the performance of the widely-used matrix probability ratio test (MSPRT) for multiple hypotheses and numerically compare its expected sample size characteristics with asymptotic bounds for these,  in a particular case  of testing of  three hypotheses about the parameter of the Bernoulli distribution.

The idea of the  MSPRT is to simultaneously run $k(k-1)/2$ SPRTs for each
pair of the hypothesized parameter  values, stopping only when all the SPRTs decide in favour of a certain hypothesis.  
Let $A_{ij}>1$ be some  constants, $1\leq i\not=j\leq k$. Then the stopping time  of the MSPRT (let us denote it  $\tau^*$) is defined as 
\begin{equation}\label{7-1}
 \min\{n\geq 1: \text{there is}\; i \;\text{such that}\; f_{\theta_i}^n(x_1,\dots,x_n)\geq A_{ij}f_{\theta_j}^n(x_1,\dots,x_n) \;\text{for all }\; j\not=i\}
\end{equation}
in which case  hypothesis $H_i$ is accepted. \cite{Armitage} showed that 
the MSPRT stops with probability one under each $H_i$, and that 
\begin{equation}\label{7-2}
 \alpha_{ij}^*\leq 1/A_{ji},\quad 1\leq i\not=j\leq k
\end{equation}
where $\alpha_{ij}^*$ is the error probability of MSPRT \eqref{7-1}.

For $k=2$ the MSPRT is an ordinary SPRT and \eqref{7-2} are the very well known Wald's inequalities for its error probabilities.

To get numerical results we consider a particular case of $k=3$ hypotheses for the parameter of success $\theta$ of the Bernoulli distribution, with $\theta_1=0.3$, $\theta_2=0.4$ and $\theta_3=0.5$.  

First of all, we will be interested in calculating the performance characteristics of the MSPRT in this particular case. 
It is easy to see that the rules of the MSPRT are based, in the Bernoulli case, on the sufficient statistics $S_n$, $n=1,2,\dots$, so the formulas of Subsection \ref{s3.2} apply for the truncated version of the MSPRT. Strictly speaking, the terminal decision at the last step, when   the MSPRT is truncated at time $N$, is not defined. But we will calculate the exact probability that MSPRT does not come  to a decision at any earlier stage, and make the probability of this so small (choosing $N$ large enough) that any concrete decision one can take in  the last step will not affect the numerical values of the error probabilities, nor those of the ESS under any one of the hypotheses.    

In \cite{Tartakovsky2014}, asymptotic formulas are obtained for the ESS of the MSPRT, so we consider this example  a good opportunity to juxtapose the really obtained  and the asymptotic values of the corresponding numerical characteristics, calculated in various practical scenarios.
We use the thresholds $A_{ji}=(k-1)/\alpha$  which make the MSPRT in \eqref{7-1} asymptotically optimal, as $\max_i\{\alpha_i\}=\alpha\to 0$ \citep[see][ Section 4.3.1]{Tartakovsky2014}.

The results of evaluations are presented in Table 1, where $\alpha_i^*$, $E_{\theta_i}\tau^*$ are the evaluated characteristics of the MSPRT, and $R_i$ the respective ratio between $E_{\theta_i}\tau^*$ and the asymptotic expression for it \citep[according to][p. 196]{Tartakovsky2014}, $i=1,2,3$.

\begin{table}[!t]
\begin{tabular}{r|rrrrrrrrr}
$\alpha$&$\alpha_1^*$&$\alpha_2^*$&$\alpha_3^*$&$E_{\theta_1}\tau^*$&$E_{\theta_2}\tau^*$&$E_{\theta_3}\tau^*$&$R_1$&$R_2$&$R_3$\\
\hline
0.1&0.026091&0.089375&0.029442&134.5&211.8&142.5&1.26&1.85&1.26\\
0.05&0.013039&0.045384&0.014829&169.4&264.9&180&1.22&1.78&1.23\\
0.025&0.006498&0.022826&0.007467&203.5&313.2&216.2&1.19&1.71&1.2\\
0.01&0.002575&0.009172&0.002981&247.4&372.4&262.7&1.16&1.63&1.16\\
0.005&0.001291&0.004596&0.001504&280&414.1&297.4&1.14&1.57&1.15\\
0.002&0.0005&0.00184&0.000594&322.8&468.9&342.8&1.12&1.52&1.13\\
0.001&0.000248&0.00092&0.000296&355.1&508.8&376.9&1.11&1.48&1.11\\
0.0005&0.000123&0.00046&0.000147&387.2&548.5&411&1.1&1.45&1.1\\
5E-07&1.14E-07&4.6E-07&1.47E-07&707.1&928.5&749.5&1.05&1.29&1.05\\
5E-09&1.1E-09&4.6E-09&1.46E-09&920.3&1175.5&975.2&1.04&1.24&1.04
\end{tabular}
\caption{ ESS:  MSPRT vs. asymptotic} 
\end{table}

\subsection{Bayes vs. MSPRT}
Now, let us numerically compare the optimal multi-hypothesis test with the MSPRT, provided both have the same levels of error probabilities $\alpha_i=\alpha$, $i=1,2,3$.
To this end, we numerically find the Lagrange multipliers 
$\lambda_i$ providing the best approximation of the error probabilities of the test
\eqref{2-9}-\eqref{2-6}  to $\alpha$, with respect to the distance
\begin{equation*}
 \max_i\{|\alpha_i(\psi,\phi)-\alpha|/\alpha\}.
\end{equation*}
The gradient-free optimization method of \cite{neldermeadarticle} works well for this fitting.
We use $\vartheta_i=\theta_i$ and $\gamma_i=1/3$, for $i=1,2,3$ as a criterion of minimization in \eqref{8-1}, i.e. we evaluate the Bayesian tests with the ``least informative'' prior distribution.
The results of fitting are presented in Table 2 (upper block).

As a competing  MSPRT we take the test \eqref{7-1}, with  $A_{ij}$ defined as $A_{ij}=A_j$ for all  $1\leq j\not=i\leq 3$, and carry out the same fitting procedure as above, with respect to $A_1,A_2,A_3$.
The results are presented in the middle block of Table 2.

In the lower block of Table 2 we placed the ratios $R_i$ between the ESS of the MSPRT ($E_{\theta_i}\tau^*$) and that of the   respective Bayesian test ($E_{\theta_i}\tau$), under each one of the hypotheses. 

The results show an astonishingly high  efficiency of the MSPRT, especially for small $\alpha$. This would not be so surprising for two hypotheses, because in this case any MSPRT is in fact an SPRT, and any Bayesian test is   an SPRT, too \citep[see][]{waldwolfowitz}, so fitting numerically both tests to given error probabilities should give a relative efficiency of about 100\%.
But we see that largely the same happens for three hypotheses, at least in the case of equal error probabilities we are examining.

The question arises whether there exist Bayesian tests ``essentially'' outperforming MSPRTs, in the case of three hypotheses. The answer is ``yes'', as the following numerical example suggests.

In a rather straightforward way, we  found a Bayesian test, corresponding to very ``unbalanced'' weights $\gamma=(0.01,0.01,0.98)$, and an MSPRT having the same error probabilities: 
$\alpha_1=0.0051$, $\alpha_2=0.089$,$\alpha_3=0.068$.
These correspond to Lagrangian multipliers 
of $\lambda_1=200$	$\lambda_2=500$
$\lambda_3=200$ for the Bayesian test and the thresholds
$\log(A_1)=	4.90$,	
$\log(A_2)=3.00$,
$\log(A_3)=1.69$ for the MSPRT, respectively.
Accordingly, we obtained
$E_{\theta_1}\tau=320.1$,	
$E_{\theta_2}\tau=258.5$,
$E_{\theta_3}\tau=101.3$ for the Bayesian test, and
$E_{\theta_1}\tau^*=139.7$,	
$E_{\theta_2}\tau^*=	239.0$,	
$E_{\theta_3}\tau^*=	134.3$ for the MSPRT.
Respectively, the weighted ESS evaluated to $C_{\gamma,\theta}(\tau)=105.07$ and							$C_{\gamma,\theta}(\tau ^*)=		135.32$, that is, nearly 29\% larger for the MSPRT in comparison with the Bayesian test.

The most desirable property an optimal test should have is that it minimizes the ESS under each one of the hypotheses, in the class of tests subject to restrictions on the error probabilities. Nevertheless, we think  this property is too strong to  be fulfilled by any sequential test, when there are  three (or more) hypotheses.
We base this opinion on the following simple observation.  Suppose there is a ``uniformly optimal'' test $\langle\phi^*,\psi^*\rangle$ in the sense that   $\alpha_{i}(\psi^*,\phi^*)=\alpha_{i}$ $i=1,\dots,k$, and for any  test $\langle\phi,\psi\rangle$ such that
$\alpha_{i}(\psi,\phi)\leq\alpha_{i}$ for $i=1,\dots,k$,  it holds
$E_{\theta_i}\tau_{\psi}\geq E_{\theta_i}\tau_{\psi^*}$ for all $i=1,\dots,k$.
Then it is obvious that, whatever be the  weights $\gamma_i\geq 0$, $i=1,\dots,k$, it holds that
$C_{\gamma,\theta}(\psi,\phi)=\sum_{i=1}^k\gamma_i E_{\theta_i}\tau_{\psi}\geq C_{\gamma,\theta}(\psi^*,\phi^*)$. 
Thus, {\em for any} set of weights $\gamma_i$, $i=1,\dots,k$ we have a test minimizing the weighted ESS under the restrictions on the error probabilities, i.e. {\em one} test 
$\langle\phi^*,\psi^*\rangle$ solves {\em all} the problems of minimization of weighted ESS we formulated in Section 2 (all those with $\vartheta=\theta$ but arbitrary $\gamma$). It seems that  this is  ``too much'' for one test when there are more than two hypotheses (it is fine for two hypotheses  because it is well known that any  Bayesian test is an SPRT). Unfortunately, the discrete nature of error probabilities in the Bernoulli model seems to be  a serious obstacle for constructing a formal counterexample in this case. We hope to be able to provide one in our future publications concerning continuous distribution families.

\begin{table}[!t]
\begin{tabular}{c|rrrrrrrr}
$\alpha$&0.1 &0.05 &0.025 &0.01 &0.005 &0.002 &0.001 &0.0005\\
\hline
$\log(\lambda_1)$&5.09&5.61&6.15&6.91&7.52&8.36&9.04&9.71\\
$\log(\lambda_2)$&5.88&6.55&7.21&8.10&8.78&9.68&10.37&11.06\\
$\log(\lambda_3)$&5.23&5.77&6.34&7.13&7.76&8.63&9.31&9.99\\
$E_{\theta_1}\tau$&113.4 &160.7 &194.4 &242.0 &276.1 &320.0 &352.6 &385.0\\
$E_{\theta_2}\tau$&136.0 &189.4 &238.4 &298.3 &340.9 &395.5 &435.7 &475.3\\
$E_{\theta_3}\tau$&115.9 &156.6 &202.4 &253.1 &289.2 &335.8 &370.4 &404.7\\
\hline
$\log(A_1)$&1.67 &2.37 &3.07 &3.96 &4.63 &5.52 &6.20 &6.88\\
$\log(A_2)$&2.81 &3.56 &4.27 &5.21 &5.90 &6.81 &7.52 &8.21\\
$\log(A_3)$&1.81 &2.50 &3.21 &4.12 &4.81 &5.72 &6.41 &7.09\\
$E_{\theta_1}\tau^*$&110.0 &153.4 &192.3 &240.1 &273.9 &317.5 &350.7 &383.1\\
$E_{\theta_2}\tau^*$&136.0 &189.4 &238.4 &298.3 &341.0 &395.0 &435.7 &475.3\\
$E_{\theta_3}\tau^*$&118.3 &163.4 &204.7 &255.1 &291.2 &337.2 &372.3 &406.7\\
\hline
$R_1$&0.970 &0.955 &0.989 &0.992 &0.993 &0.992 &0.995 &0.995\\
$R_2$&1.000 &1.000 &1.000 &1.000 &1.000 &0.999 &1.000 &1.000\\
$R_3$&1.010 &1.043 &1.011 &1.007 &1.008 &1.004 &1.005 &1.005\\
\end{tabular}
\caption{ Relative efficiency of the MSPRT with respect to the Bayesian test} 
\end{table}
\subsection {The Kiefer-Weiss problem for multi-hypothesis testing}
In this subsection we propose a construction of a test which might be helpful for solution  of the Kiefer-Weiss problem for multiple hypotheses and present a numerical  example where the proposed test provides an approximate solution to the Kiefer-Weiss problem in the case of three hypotheses about the parameter of the Bernoulli model.

Let $\theta_1<\theta_2<\dots<\theta_K$ be the hypothezised parameter values, $K\geq 2$. Generalizing  the Kiefer-Weiss problem from the case of $K=2$ hypotheses \citep[see][]{Kiefer} let us say that the Kiefer-Weiss problem  for $K\geq 2$ hypotheses is to find a sequential test $\langle \psi,\phi\rangle$ which minimizes $\sup_{\theta\in(\theta_1,\theta_2)} E_\theta\tau_\psi$ in the class of tests subject to restrictions on the error probabilities \eqref{1}.  

\cite{Kiefer} and \cite{Weiss} noted that in some symmetrical cases the solution can be obtained  as a solution to a much simpler problem (called modified Kiefer-Weiss problem nowadays). This latter problem is to find a test minimizing $E_{\vartheta_1} \tau_\psi$ among the tests satisfying the restrictions on the error probabilities, where $\vartheta_1$ is  some point in $(\theta_1, \theta_2)$.

For the general multi-hypothesis case we propose the following generalization of this construction. Let $\vartheta_i\in(\theta_i, \theta_{i+1})$, for $i=1,2,\dots,k-1$, be some parameter points. And let $\gamma_i\in[0,1]$, $i=1,2,\dots,k-1$, be some weights (such that $\sum_{i=1}^{k-1}\gamma_i=1$). Recall that
\begin{equation}
 C_{\gamma,\vartheta}(\psi)=\sum_{i=1}^{k-1}\gamma_i E_{\vartheta_i}\tau_{\psi}
\end{equation}

\begin{proposition}\label{p4-1}
Let us suppose there is a test $\langle\psi^*,\phi^*\rangle$,
with some $\vartheta_i\in(\theta_i, \theta_{i+1})$, and $\gamma_i\geq 0$, $i=1,2,\dots,k-1$,  $\sum_{i=1}^{k-1}\gamma_i=1$, such that
\begin{equation}\label{10-1}
 C_{\gamma,\vartheta}(\psi^*)+\sum_{i\not =j}\lambda_{ij}\alpha_{ij}(\psi^*,\phi^*)\leq C_{\gamma,\vartheta}(\psi)+\sum_{i\not =j}\lambda_{ij}\alpha_{ij}(\psi,\phi)
\end{equation}
for all sequential tests $\langle\psi,\phi\rangle$, and that
\begin{equation}\label{10-4}
 \alpha_{ij}(\psi^*,\phi^*)=\alpha_{ij}, \;\mbox{for all}\;1\leq i\not=j\leq k.
\end{equation}
Additionally, let us suppose that
\begin{equation}\label{10-5}
E_{\vartheta_i}\tau_{\psi^*}=\sup_{\theta\in(\theta_1,\theta_{k})} E_{\theta}\tau_{\psi^*}\;\mbox{for all}\;1\leq i\leq k-1.
\end{equation}

Then for any sequential test $\langle\psi,\phi\rangle$ satisfying
\begin{equation}\label{10-2}
 \alpha_{ij}(\psi,\phi)\leq\alpha_{ij}, \;\mbox{for all}\quad 1\leq i\not=j\leq k,
\end{equation}
it holds
\begin{equation}\label{10-6}
 \sup_{\theta\in(\theta_1,\theta_k)}E_{\theta}\tau_{\psi^*}\leq \sup_{\theta\in(\theta_1,\theta_k)}E_{\theta}\tau_{\psi} ,
\end{equation}
i.e. $\langle\psi^*,\phi^*\rangle$ solves the Kiefer-Weiss problem.

\end{proposition}
% Let us call the problem of minimization of  $\sum _{i=1}^{k-1} \gamma_{i} E_{\vartheta_i}\tau_\psi$ among all the tests satisfying \eqref{}, modified multi-hypothesis Kiefer-Weiss problem.
{\bf Proof.} It follows from \eqref{10-1}, \eqref{10-4} and \eqref{10-2} that
\begin{eqnarray*}\label{10-3}
C_{\gamma,\vartheta}(\psi^*)+\sum_{i\not =j}\lambda_{ij}\alpha_{ij}= C_{\gamma,\vartheta}(\psi^*)+\sum_{i\not =j}\lambda_{ij}\alpha_{ij}(\psi^*,\phi^*)\\
\leq C_{\gamma,\vartheta}(\psi)+\sum_{i\not =j}\lambda_{ij}\alpha_{ij}(\psi,\phi)\leq C_{\gamma,\vartheta}(\psi)+\sum_{i\not =j}\lambda_{ij}\alpha_{ij}
\end{eqnarray*}
for any test $\langle\psi,\phi\rangle$ satisfying \eqref{10-2}, so
$$
C_{\gamma,\vartheta}(\psi^*)=\sum_{i=1}^{k-1}\gamma_i E_{\vartheta_i}\tau_{\psi^*}\leq C_{\gamma,\vartheta}(\psi)=\sum_{i=1}^{k-1}\gamma_i E_{\vartheta_i}\tau_{\psi}\leq \sup_{\theta\in(\theta_1,\theta_k)}E_{\theta}\tau_{\psi}.
$$
But, due to \eqref{10-5},
\begin{equation*}
 \sum_{i=1}^{k-1}\gamma_i E_{\vartheta_i}\tau_{\psi^*=}\sup_{\theta\in(\theta_1,\theta_{k})} E_{\theta}\tau_{\psi^*},
\end{equation*}
thus \eqref{10-6} follows. $\Box$
\begin{remark} The modification of Proposition \ref{p4-1} to be used with restrictions on $\alpha_i$ rather than on $\alpha_{ij}$ is straightforward: just using $\lambda_i, \alpha_i$ instead of $\lambda_{ij}$ and $\alpha_{ij} $, respectively.
\end{remark}
\begin{remark}
 We conjecture that, when sampling from exponential families of distributions, the tests constructed in Proposition \ref{p4-1} for multiple hypotheses (even without condition \eqref{10-5}), are always truncated, just like those in the modified Kiefer-Weiss problem  for two hypotheses are, when $\vartheta_1\in(\theta_1,\theta_2)$. Using our program in \cite{multihypothesisGit} it is easy to see this for any number of hypotheses in the Bernoulli case.
\end{remark}
\begin{remark}
 Proposition \ref{p4-1} is valid for any number of hypotheses for any parametric family of distributions.
\end{remark}

Let us consider now an example of of a numerical solution to the Kiefer-Weiss problem for Bernoulli model, in the case of three hypotheses.

Let $\theta_1=0.3$, $\theta_2=0.5$ and $\theta_3=0.7$.  
We took $N=1200$, $\gamma_1=\gamma_2=0.5$ and $\lambda_1=\lambda_2=\lambda_3=200$ and used the function {\tt OptTest} from the program code in  \cite{multihypothesisGit}
to produce tests satisfying condition \eqref{10-1} (minimizing the Lagrangian function). 
 To comply with  \eqref{10-5}, after a simple numerical optimization over $\vartheta_1=1-\vartheta_2$ we found that 
for $\vartheta_1=0.4026$, $\vartheta_2=0.5974$  it holds
$$\max_{\theta\in[0.3,0.7]} E_{\theta}\tau_{\psi^*}=56.2=E_{\vartheta_1}\tau_{\psi^*}=E_{\vartheta_2}\tau_{\psi^*}$$
To calculate the error probabilities we used the function {\tt PAccept} in \cite{multihypothesisGit}, and obtained
$\alpha_1(\psi^ *,\phi^*)=\alpha_3(\psi^*,\phi^*)=0.037$ and $\alpha_2(\psi^ *,\phi^*)=0.07$.
Thus, we have a numerical solution of the Kiefer-Weiss problem under restrictions $\alpha_1=\alpha_3=0.037$ and $\alpha_2=0.07$. The optimal test is truncated at $N=160$. The function {\tt maxNumber} can be used to see the maximum number of steps a test requires.

To compare the Kiefer-Weiss solution with a Bayesian test we used the same function {\tt OptTest}, now with $\theta_i=\vartheta_i, i=1,2,3$ and $\gamma_i=1/3$, $i=1,2,3$ at the truncation level $N=1200$ using the Nelder-Mead optimization to get (as close as possible to $\alpha_1=\alpha_3=0.037$ and $\alpha_2=0.07$). The fitted values are $\alpha_1=\alpha_3=0.0370$ and $\alpha_2=0.0704$ and the maximum ESS of 60.2. Thus, the Kiefer-Weiss solution saves about 10\% of observations, on the average, in comparison with the optimal Bayesian test.

\section{Conclusions and further work}
In this paper, we proposed a computer-oriented  method of construction of  sequential  multi-hypothesis tests, minimizing a weighted expected sample number (ESS).

For the particular case of sampling from a Bernoulli population, we developed a computational scheme for evaluating the optimal tests and calculating the numerical characteristics of sequential tests based on  sufficient statistics. An implementation of the algorithms in the R programming language has been published in a GitHub repository  \cite{multihypothesisGit}. 

A numerical evaluation of the widely-used multi-hypothesis sequential probability ratio test is carried out for the case of three simple hypotheses about the parameter of the Bernoulli distribution, and a numerical comparison is made with the asymptotic expressions for the ESS of the asymptotically optimal MSPRT.

For a series of error probabilities %$\alpha_1=\alpha_2=\alpha_3=\alpha$ 
we evaluated the ESS of the Bayesian test and compared it with that of the MSPRT having the same error probabilities, in which case the MSPRT exhibited a very high efficiency. On the other hand, we found a numerical example where the MSPRT is substantially less efficient than the optimal Bayesian test.

We proposed a method of numerical solution of the multi-hypothesis Kiefer-Weiss problem. The proposed method is applied to three-hypothesis Kiefer-Weiss problem for the Bernoulli. Numerial results are given.

A very immediate extension of this work could be developing computational algorithms for construction and performance evaluation of optimal sequential  multi-hypothesis tests for other parametric families, first of all for one-parameter exponential families \citep[cf. ][]{novikovfarkhshatovDiscrete}.

The method we applied in this paper for i.i.d. observations can in fact  be used for much more general models. For example, it can be applied to the models considered in \cite{Liu},   where numerical methods of performance evaluation of the MSPRT for non-i.i.d. observations are developed. It would be interesting to carry out  a comparison study between the MSPRT and our optimal tests. Extensions to models with dependent observations are also  possible.

%The case of non-exponential families can also be treated numerically
The proposed method for solution of the Kiefer-Weiss problem can be extended to other parametric families.

Another expected application  is an extension of sequentially planned tests  for two hypotheses \citep{spprt} to the case of multiple hypotheses 
\citep{Novikov2009Kybernetika}.

\section*{Acknowledgements}
The author gratefully acknowledges a partial support of the National Researchers System (SNI) by CONACyT, Mexico, for this work. 

The author thanks  the anonymous Reviewers and the Associate Editor for very substantial comments and suggestions on improving  earlier versions of this work. 
\section*{Appendix. Proof of (\ref{9-1})}
Let us define $\alpha_{ij}(n,\phi)$ as the error probability of the fixed-sample-size test 
 based on $n$ observations and using the decision rule from \eqref{2-9}. 
It follows from Theorem 3 in \cite{novikovMultiple} that 
$$
\int v_nd\mu^n=\sum_{i\not=j}\lambda_{ij} \alpha_{ij}(n,\phi),
$$ 

Let us prove that for any $i\not=j$ such that $\lambda_{ij}>0$ $\alpha_{ij}(n, \phi)\to 0$, as $n\to\infty$.

We have
$$
\alpha_{ij}(n,\phi)= P_{\theta_i}(\sum_{l:l\not=j}\lambda_{lj} f_{\theta_l}^n=v_n)\leq P_{\theta_i}(\sum_{l:l\not=j}\lambda_{lj}f_{\theta_l}^n\leq\sum_{l:l\not=i}\lambda_{li}f_{\theta_l}^n)
$$
$$\leq P_{\theta_i}(\sum_{l:l\not=j}\lambda_{lj}\frac{f_{\theta_l}^n}{f_{\theta_i}^n}\leq\sum_{l:l\not=i}\lambda_{li}\frac{f_{\theta_l}^n}{f_{\theta_i}^n})\leq P_{\theta_i}( \sum_{l:l\not=j}\lambda_{lj}\frac{f_{\theta_l}^n}{f_{\theta_i}^n}\leq\sum_{l:l\not=i}\lambda_{li}\frac{f_{\theta_l}^n}{f_{\theta_i}^n}) 
$$
$$
\leq P_{\theta_i}( \lambda_{ij}+\sum_{l:l\not=i,j}\lambda_{lj}\frac{f_{\theta_l}^n}{f_{\theta_i}^n}\leq\sum_{l:l\not=i}\lambda_{li}\frac{f_{\theta_l}^n }{f_{\theta_i}^n}) \to 0, \quad \mbox{as}\quad n\to\infty.
$$
This latter holds because

$$
\frac{f_{\theta_l}^n }{f_{\theta_i}^n}\to 0, \;\mbox{as}\;n\to\infty,
$$
in $P_{\theta_i}$-probability for any 
$l\not= i$. Indeed, by the Markov inequality
$$
P_{\theta_i}({\frac{f_{\theta_l}^n }{f_{\theta_i}^n}}>\epsilon)= P_{\theta_i}(\sqrt{\frac{f_{\theta_l}^n }{f_{\theta_i}^n}}>\sqrt{\epsilon}) 
\leq E_{\theta_i}\sqrt{\frac{f_{\theta_l}^n }{f_{\theta_i}^n}}/\sqrt{\epsilon}$$
$$=\left(\int \left(f_{\theta_i}f_{\theta_l}\right)^{1/2}d\mu\right)^n/\sqrt{\epsilon}\to 0,\quad\mbox{as}\quad n\to\infty,
$$
because $\int \left(f_{\theta_i}f_{\theta_l}\right)^{1/2}d\mu <1$ for $l\not=i$, due to the Cauchy-Schwarz inequality.


\begin{thebibliography}{19}
\newcommand{\enquote}[1]{``#1''}
\providecommand{\natexlab}[1]{#1}
\providecommand{\url}[1]{\normalfont{#1}}
\providecommand{\urlprefix}{}

\bibitem[Armitage(1950)]{Armitage}
Armitage, P. 1950. ``Sequential analysis with more than two alternative
  hypotheses, and its relation to discriminant function analysis.''
  \emph{Journal of the Royal Statistical Society B} 12: 137--144.

\bibitem[Baum and Veeravalli(1994)]{Baum}
Baum, C.~W., and V.~V. Veeravalli. 1994. ``A Sequential Procedure for
  Multihypothesis Testing.'' \emph{{IEEE} Transactions on Information Theory}
  40 (6): 1994--2007.

\bibitem[Blackwell and Girshick(1954)]{blackwell}
Blackwell, D., and M.~A. Girshick. 1954. \emph{Theory of games and statistical
  decisions}. John Wiley and Sons, Inc.

\bibitem[Chow, Robbins, and Siegmund(1971)]{Chow}
Chow, Y.S, H.~Robbins, and S.~Siegmund. 1971. \emph{Great Expectations: The
  Theory of Optimal Stopping}. Houghton Mifflin.

\bibitem[Eales and Jennison(1992)]{eales}
Eales, J.~D., and C.~Jennison. 1992. ``{An improved method for deriving optimal
  one-sided group sequential tests}.'' \emph{Biometrika} 79 (1): 13--24.
  \urlprefix\url{https://doi.org/10.1093/biomet/79.1.13}.

\bibitem[Kiefer and Weiss(1957)]{Kiefer}
Kiefer, J., and L.~Weiss. 1957. ``Some properties of generalized sequential
  probability ratio tests.'' \emph{Annals of Mathematical Statistics} 28:
  57--75.

\bibitem[Liu, Gao, and Li(2016)]{Liu}
Liu, Y., Y.~Gao, and X.~Rong Li. 2016. ``Operating Characteristic and Average
  Sample Number of Binary and Multi-Hypothesis Sequential Probability Ratio
  Test.'' \emph{IEEE Transactions on Signal Processing} 64 (12): 3167--3179.

\bibitem[Lorden(1980)]{Lorden}
Lorden, G. 1980. ``Structure of sequential tests minimizing an expected sample
  size.'' \emph{Zeitschrift f\"ur Wahrscheinlichkeitstheorie und Verwandte
  Gebiete} 51 (3): 291--302.

\bibitem[Nelder and Mead(1965)]{neldermeadarticle}
Nelder, J.~A., and T.~Mead. 1965. ``A simplex method for function
  minimization.'' \emph{Computer Journal} 7 (4): 308–313.

\bibitem[Novikov(2009{\natexlab{a}})]{Novikov2009Kybernetika}
Novikov, A. 2009{\natexlab{a}}. ``Optimal Sequential Multiple Hypothesis
  Testing in Presence of Control Variables.'' \emph{Kybernetika} 45 (3):
  507--528.

\bibitem[Novikov(2009{\natexlab{b}})]{novikovMultiple}
Novikov, A. 2009{\natexlab{b}}. ``Optimal Sequential Multiple Hypothesis
  Tests.'' \emph{Kybernetika} 45 (2): 309--330.

\bibitem[Novikov(2022)]{spprt}
Novikov, A. 2022. ``Optimal design and performance evaluation of sequentially
  planned hypothesis tests.''
  \urlprefix\url{https://arxiv.org/abs/2210.07203}.

\bibitem[Novikov(2023)]{multihypothesisGit}
Novikov, A. 2023. ``{ An R Project for Construction and Performance Evaluation
  of Sequential Multi-Hypothesis Tests}.''
  \url{https://github.com/HOBuKOB-MEX/multihypothesis}.

\bibitem[Novikov and Farkhshatov(2022)]{novikovfarkhshatovDiscrete}
Novikov, A., and F.~Farkhshatov. 2022. ``Design and performance evaluation in
  Kiefer-Weiss problems when sampling from discrete exponential families.''
  \emph{Sequential Analysis} 41 (04): 417 -- 434.

\bibitem[{R Core Team}(2013)]{R}
{R Core Team}. 2013. \emph{R: A Language and Environment for Statistical
  Computing}. Vienna, Austria: R Foundation for Statistical Computing.
  \urlprefix\url{http://www.R-project.org/}.

\bibitem[Shiryaev(1978)]{Shiryaev}
Shiryaev, A.~N. 1978. \emph{Optimal stopping rules}. Berlin: Springer.

\bibitem[Tartakovsky, Nikiforov, and Basseville(2015)]{Tartakovsky2014}
Tartakovsky, A.~G., I.~V. Nikiforov, and M.~Basseville. 2015. \emph{Sequential
  analysis: hypothesis testing and changepoint detection}. Boca Raton, Florida:
  Chapman \& Hall/CRC Press.

\bibitem[Wald and Wolfowitz(1948)]{waldwolfowitz}
Wald, A., and J.~Wolfowitz. 1948. ``Optimum character of the sequential
  probability ratio test.'' \emph{Annals of Mathematical Statistics} 19 (3):
  326--339.

\bibitem[Weiss(1962)]{Weiss}
Weiss, L. 1962. ``On sequential tests which minimize the maximum expected
  sample size.'' \emph{Journal of American Statistical Assocciation} 57:
  551--566.

\end{thebibliography}
\end{document}